\documentclass[aps,10pt,twocolumn,amsmath,amssymb,groupedaddress,prl,floatfix]{revtex4-2}

\usepackage{graphicx} 
\usepackage{subfigure}
\usepackage{xcolor}
\usepackage[integrals]{wasysym}
\usepackage{dcolumn}
\usepackage{epstopdf}
\usepackage[]{hyperref}
\usepackage{textcomp} 
\usepackage{gensymb}
\usepackage{soul} 
\usepackage{siunitx}
\usepackage{amsmath}
\usepackage{amssymb}

\begin{document}

\title{Curving Origami with Mechanical Frustration}

\author{T. Jules\textsuperscript{1,2}}
\email{theo.jules@ens-lyon.fr}
\author{F. Lechenault\textsuperscript{2}}
\author{M. Adda-Bedia\textsuperscript{1}}

\affiliation{\textsuperscript{1}Universit\'e de Lyon, Ecole Normale Sup\'erieure de Lyon, Universit\'e Claude Bernard, CNRS, Laboratoire de Physique, F-69342 Lyon, France}
\affiliation{\textsuperscript{2}Laboratoire de Physique de l'Ecole Normale Sup\'erieure, ENS, PSL Research University, CNRS, Sorbonne University, Universit\'e Paris Diderot, Sorbonne Paris Cit\'e, 75005 Paris, France}

\date{\today}

\begin{abstract}

We study the three-dimensional equilibrium shape of a shell formed by a deployed accordion-like origami, made from an elastic sheet decorated by a series of parallel creases crossed by a central longitudinal crease. Surprisingly, while the imprinted crease network does not exhibit a geodesic curvature, the emergent structure is characterized by an effective curvature produced by the deformed central fold. Moreover, both finite element analysis and manually made mylar origamis show a robust empirical relation between the imprinted crease network's dimensions and the apparent curvature. A detailed examination of this geometrical relation shows the existence of three typical elastic deformations, which in turn induce three distinct types of morphogenesis. We characterize the corresponding kinematics of crease network deformations and determine their phase diagram. Taking advantage of the frustration caused by the competition between crease stiffness and kinematics of crease network deformations, we provide a novel tool for designing curved origami structures constrained by strong geometrical properties.

\end{abstract}

\maketitle

\section{Introduction}

Origamis are the three-dimensional structures obtained by folding a thin sheet following a specific imprinted pattern of creases. Through their apparent scalability and the infinite number of crease network combinations, origamis offer new methods to produce mechanical metamaterials~\cite{Lv2014, Silverberg2014, Overvelde2016}. Their innovation potential is only limited to our understanding of their kinematics of deformation and mechanical properties. Interestingly, even a ``simple'' mathematical model such as rigid-foldable origamis~\cite{Tachi2010}, where the geometry imposes heavy, kinetic constraints through infinitely stiff faces, already yields exciting behavior such as non-zero Gaussian curvature, saddle-shaped configuration, and auxetic deformation~\cite{Schenk2013, Wei2013, Pratapa2019}. This construction leads to inverse origami design where a programmed tessellation pattern produces a target deployed shape~\cite{Dudte2016, Nassar2017,  Hayakawa2020}.

However, both initial and target structures are only a subset of a larger configurational space, taking into account the elastic deformation of the faces and the creases' stiffness. Both properties are crucial in understanding folded elastic sheets' behavior. Indeed, the freedom granted by the bending and stretching of the faces enables continuous deformations between stable configurations in multistable structures, otherwise forbidden with rigid faces constraints~\cite{Reid2017, Silverberg2015}. Still, these considerations rely on fine-tuning the pattern of creases to obtain, at rest, a predefined deployed shape with flat faces.

In contrast, even crumpled paper produces a complex three-dimensional structure through a random pattern of creases~\cite{Sultan2006, Gottesman2018} and the local competition between different modes of deformation~\cite{Amar1997}. The internal frustration generated by folding imposes local mechanical equilibrium that shapes the origami. This behavior is evident for a simple vertex formed by intersecting creases~\cite{Walker2018, Andrade-Silva2019} or in the saddle-like shape obtained when the imprinted crease displays geodesic curvature~\cite{Demaine2011, Dias2012, Mouthuy2012}.

In this letter, we explore the possibility of producing curvature, an essential element in shaping origami structures, without starting from curved-crease sculpture~\cite{Demaine2011, Dias2012} nor intricate crease network as in Resch patterns~\cite{Tachi2013a, Chen2016}. To illustrate our proposal, we study a folding pattern, coined the curved-accordion, and exploit the frustration-induced deployment to display an apparent curvature along a single long crease. We describe the typical deformations of the resulting three-dimensional structure. We show that both its local and global shape are controlled by the crease network's geometry and elasticity. This intriguing result confirms that, despite the importance of the elastic response of the underlying material, the crease network structure acts as a backbone for the origami, similarly to elastic gridshells~\cite{Baek2019}.

\section{Materials and methods}

\subsection{Folding pattern and experiments}

\begin{figure}[tb]
\begin{center}
\includegraphics[width=\linewidth]{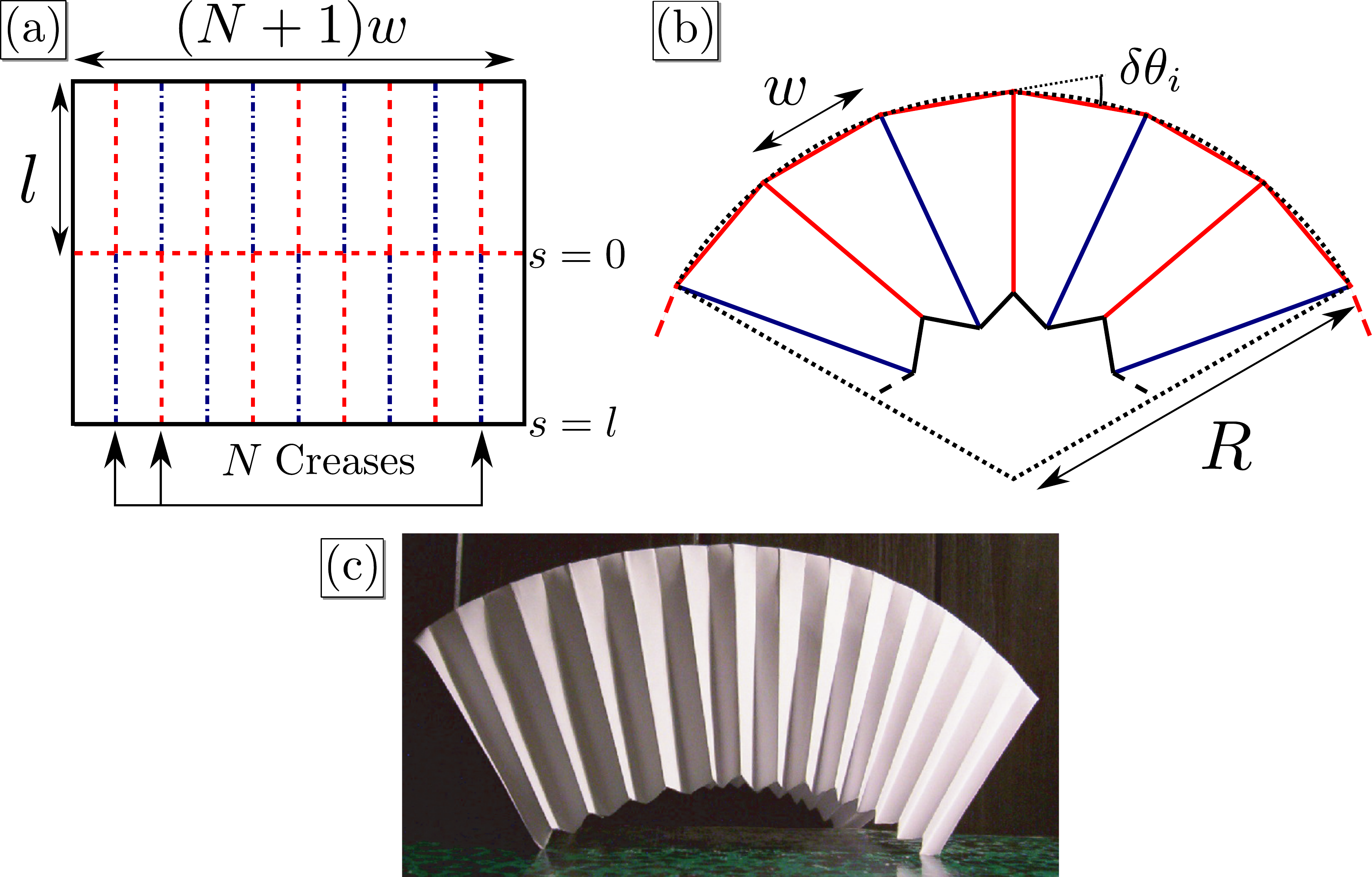}\\
\caption{Schematics of the curved accordion origami. (a) The imprinted crease network. Each face is a rectangle of dimensions $l\times w$. The red (resp. blue) dashed (resp. dash-dotted) lines represent the mountain (resp. valley) folds. (b) Resulting structure after folding and deployment. $\delta\theta_i$ is the angle between successive segments of the central crease intersecting at the $i^{\mathrm{th}}$ vertex. $R$ is the apparent radius of curvature along the central crease. (c) Curved accordion origami made from A4--paper sheet with $N=27$. }
\label{fig:OrigamiShape}
\end{center}
\end{figure}

The folding pattern for the curved-accordion origami is depicted in Fig.~\ref{fig:OrigamiShape}(a). A rectangular sheet of thickness $t$ is decorated by a single central mountain crease crossed by $N$ equally spaced perpendicular transverse creases. The latter have alternating directions that switch when crossing the central crease. The length $l$ and width $w$ of every resulting rectangular face are identical. In other words, the proposed pattern corresponds to the extreme case of Miura folding~\cite{Schenk2013} with a single straight all-mountain crease and right angles between folds.

To study the relation between the shape of the origami and the pattern of folds, we manually folded curved-accordions from 100~$\mu$m thick polyethylene terephthalate (mylar) flat sheets with a various number of creases (\SIrange{9}{22}{}), length $l$ (\SIrange{10}{140}{\milli\meter}), and width $w$ (\SIrange{9}{20}{\milli\meter}). Contrary to paper sheets, where the damage produced during the folding process yields unknown crease properties~\cite{Grey2020}, mylar sheets generate folds with a reproducible stiffness based on the elastic properties of the material~\cite{Jules2019}. After initial folding, we deployed each origami by pulling on the faces manually to set the rest angle $\Psi_0$ at approximately $90\degree$. As we do not have a precise method to measure it, each rest angle's exact value is unknown. We obtained a three-dimensional shell structure at equilibrium (see Fig.~\ref{fig:OrigamiShape}(c)) that is stable and rigid in the direction perpendicular to the central crease, similar to the rigidity displayed by an elastic spherical shell~\cite{Lazarus2012}. To investigate its shape, we let the origami lay on the free edges and took photos of the side such that the central crease is in the camera's focal plane. Alternately, we used structured light scanning to analyze the three-dimensional structures (See Supplemntary Materials).

\subsection{Numerical simulations}

In parallel, we simulated the folding of an initially flat sheet following the pattern described in Fig.~\ref{fig:OrigamiShape}(a) using finite element methods (FEM) with the software COMSOL. To do so, we followed the phase-field blueprint used in~\cite{Andrade-Silva2019}. We exploited the software's multi-physics capabilities to reproduce the change of reference configuration during folding with local thermal expansions. This protocol was experimentally tested to produce self-folding sheets~\cite{Na2014}. We chose unique elastic properties for the whole sheet, with a Young modulus $E=4$ GPa and a Poisson ratio $\nu=0.4$, based on previous work with Mylar sheets~\cite{Jules2019} and tabulated values, while setting a thickness identical to the experiments. The creases characteristic width $S_0 = 3t$ is chosen as a typical size for a manually folded crease~\cite{Jules2019, Benusiglio2012}, while the heating temperature and thermal expansion coefficient are selected to impose a unique rest angle $\Psi_0=\pi/2$ for all creases. These parameters are enough to define the faces' stretching and bending stiffness, as well as the creases' stiffness~\cite{Jules2019}. We only considered an odd number of transverse creases $N$ to take advantage of the symmetry at the crease $(N+1)/2$ and reduce the computational load. We simulated accordions for a wide range of dimensions, for $l$ from \SIrange{5}{200}{\milli\meter}, for $w$ from \SIrange{3.2}{40}{\milli\meter}, and with N from 3 to 31. We describe more details on the simulations protocol in Supplemntary Materials. 

\section{Results}

\subsection{Curvature of the origami}

\begin{figure}[tb]
\begin{center}
\includegraphics[width=0.8\linewidth]{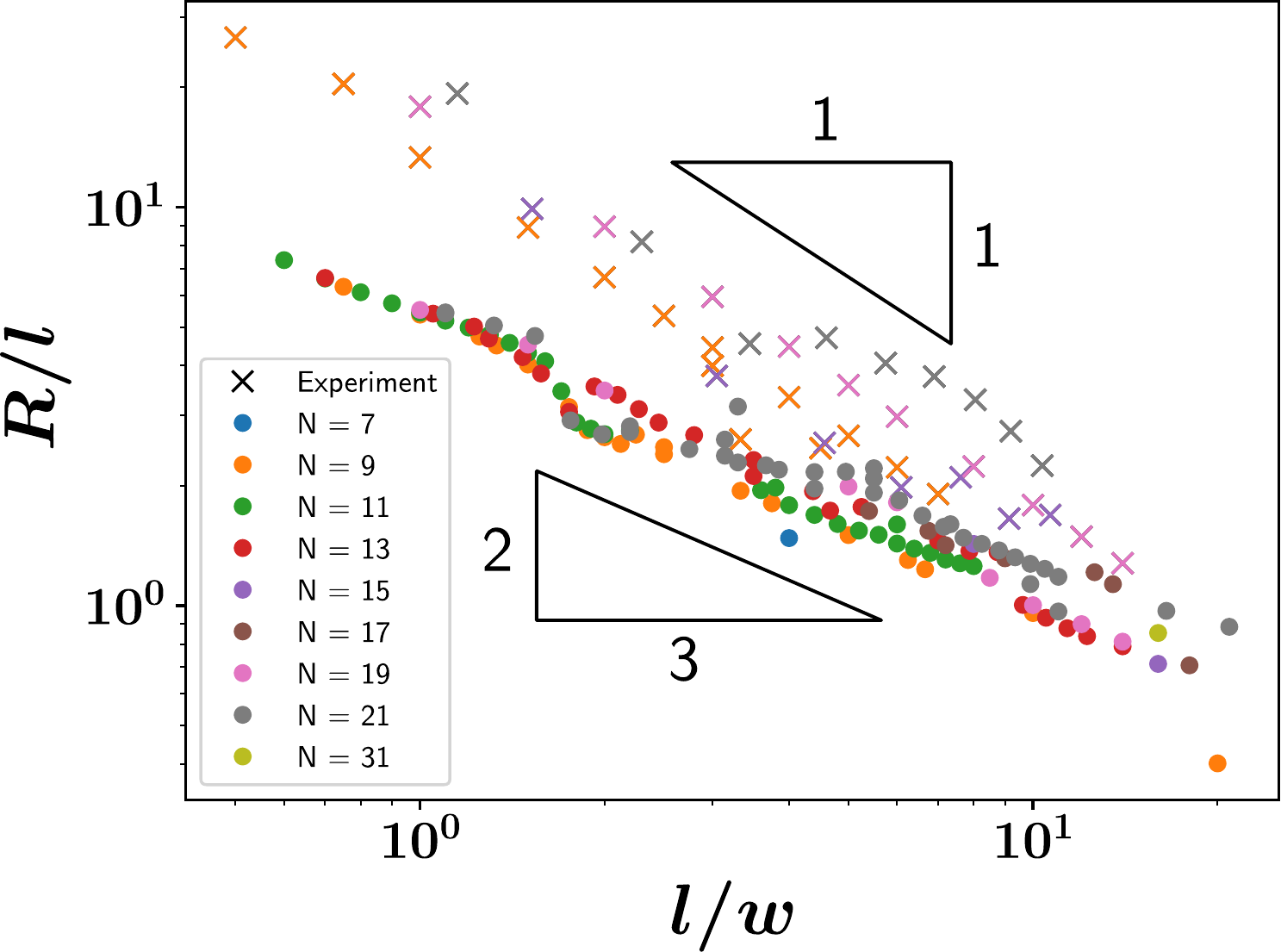}
\caption{Evolution of normalized radius $R/l$ for both simulations (full circles) and experiments (cross) with respect to the faces' aspect ratio. Each color corresponds to a given number of transverse creases $N$. For the experiments, we measured $R$ with $N=9$, 15, 19, and 22.}
\label{fig:RcEvol}
\end{center}
\end{figure}

For both experimental and simulated origamis, the central crease naturally curves, as evidenced in Fig.~\ref{fig:OrigamiShape}(c). In fact, the transverse creases divide the central one into multiple straight segments delimited by the vertices. The mechanical equilibrium resulting from the competition between rigid creases connected by flexible faces requires the $i$th vertex to exhibit an angle $\delta\theta_i$ between the $i$th and $(i+1)$th segment~\cite{Lechenault2015}. The contribution of all the $\delta\theta_i$ yields a discrete curvature
\begin{align}
\kappa = \frac{1}{R} = \frac{\sum_{i=1}^N \delta\theta_i}{(N+1)w}\;.
\label{eq:DiscCurv}
\end{align}
To recover the corresponding radius of curvature $R$, we measured the radius of the best fitting arc to the projection of the central crease on a plane, as shown in Fig.~\ref{fig:OrigamiShape}(b). For the experimental structures, the plane of projection is defined by the image. We used a singular value decomposition algorithm to find the closest plane to the three-dimensional coordinates of all the mesh points from the central crease for the simulated origamis.

Fig.~\ref{fig:RcEvol} shows slightly different behavior between the experiments and the simulations. For moderately large number of creases ($N\geq 7$), the latter displays a clear empirical non-linear relation between $R$ and the dimensions of the faces 
\begin{align}
R_\text{simu} \propto w^{2/3}l^{1/3}\;.
\label{eq:RcFacesSimu}
\end{align} 
In contrast, the experimental data seem to follow a linear relation 
\begin{align}
R_\text{exp} \propto w\;,
\label{eq:RcFacesExp}
\end{align}
which is a direct consequence of Eq.~(\ref{eq:DiscCurv}) when assuming the angles $\delta\theta_i$ do not depend on the facet size. This points to a decoupling of the vertices in the explored experimental regime, which might not be achieved in the simulations.

Overall, the simulations overestimate the curvature of the experimental accordions. It showcases two significant differences between our naive experimental protocol and FEM: the precise control of every rest angle and non-elastic mechanical behavior. On the one hand, the angles $\delta\theta_i$ highly depend on the value of the rest angle at each vertex~\cite{Andrade-Silva2019}. Indeed, we observed a non-linear increase of $R$ with respect to the rest angle during the folding of the accordion in the simulations  (See Supplemntary Materials). On the other hand, the folding pattern requires a stress concentration near the vertices, where the geometric incompatibility is the strongest. Consequently, the rest angle of the transverse crease is larger in this region than near the free edges.

\subsection{Deformation modes}

\begin{figure}[htb]
\begin{center}
\includegraphics[width=0.8\linewidth]{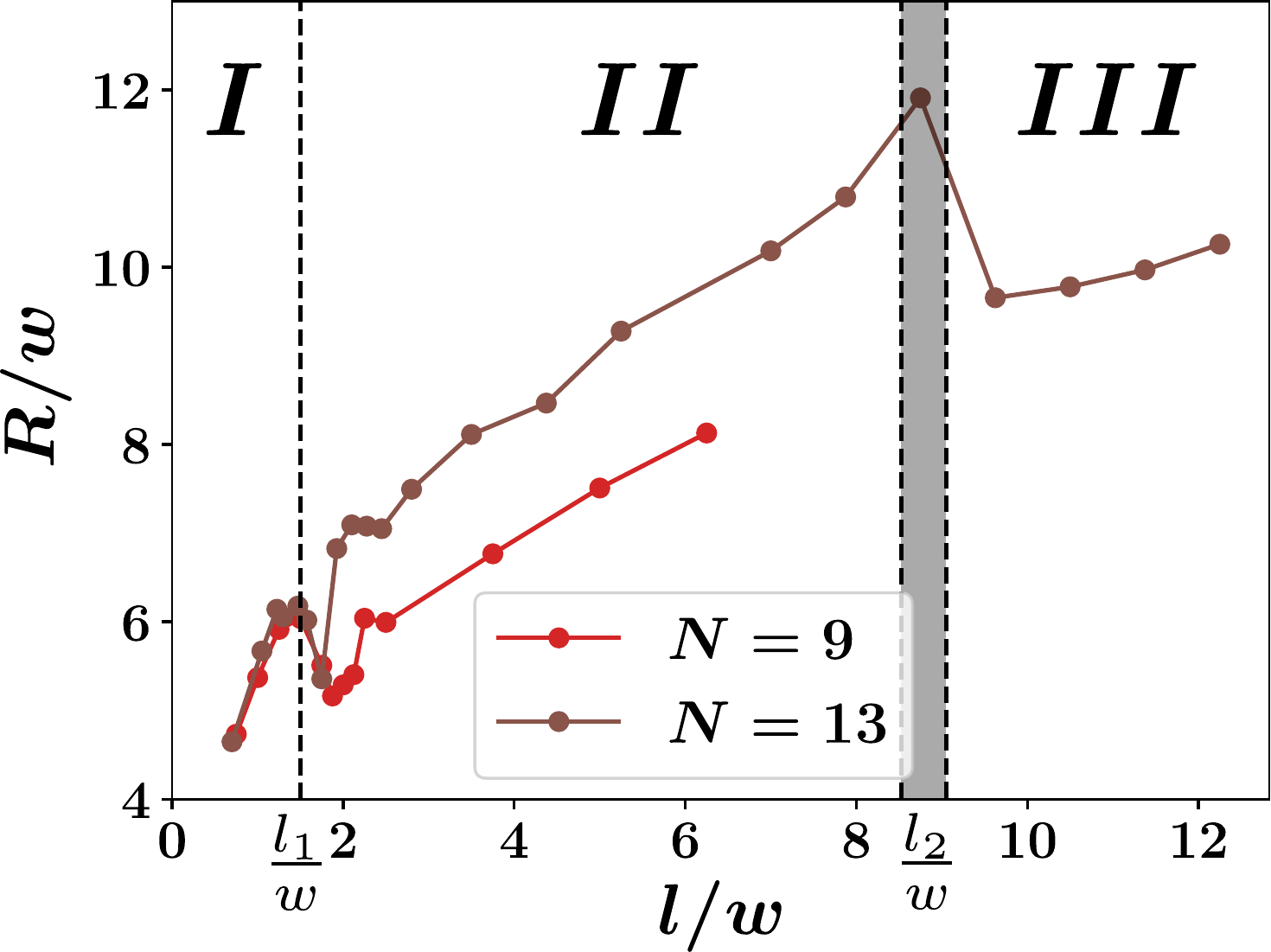}
\caption{Non monotonous evolution of $R$ with respect to the face length $l$ for $w=14.3$~mm and $N=9$ and 13. The dotted vertical lines indicate the approximate transition between three typical types of elastic deformation given by $l_1\approx 1.5 w$ and $l_2 \approx 9w$. The point in the grey region displays local deformations typical of both region $II$ and region $III$.}
\label{fig:NonMonoRc}
\end{center}
\end{figure}

\begin{figure}[htb]
\begin{center}
\includegraphics[width=\linewidth]{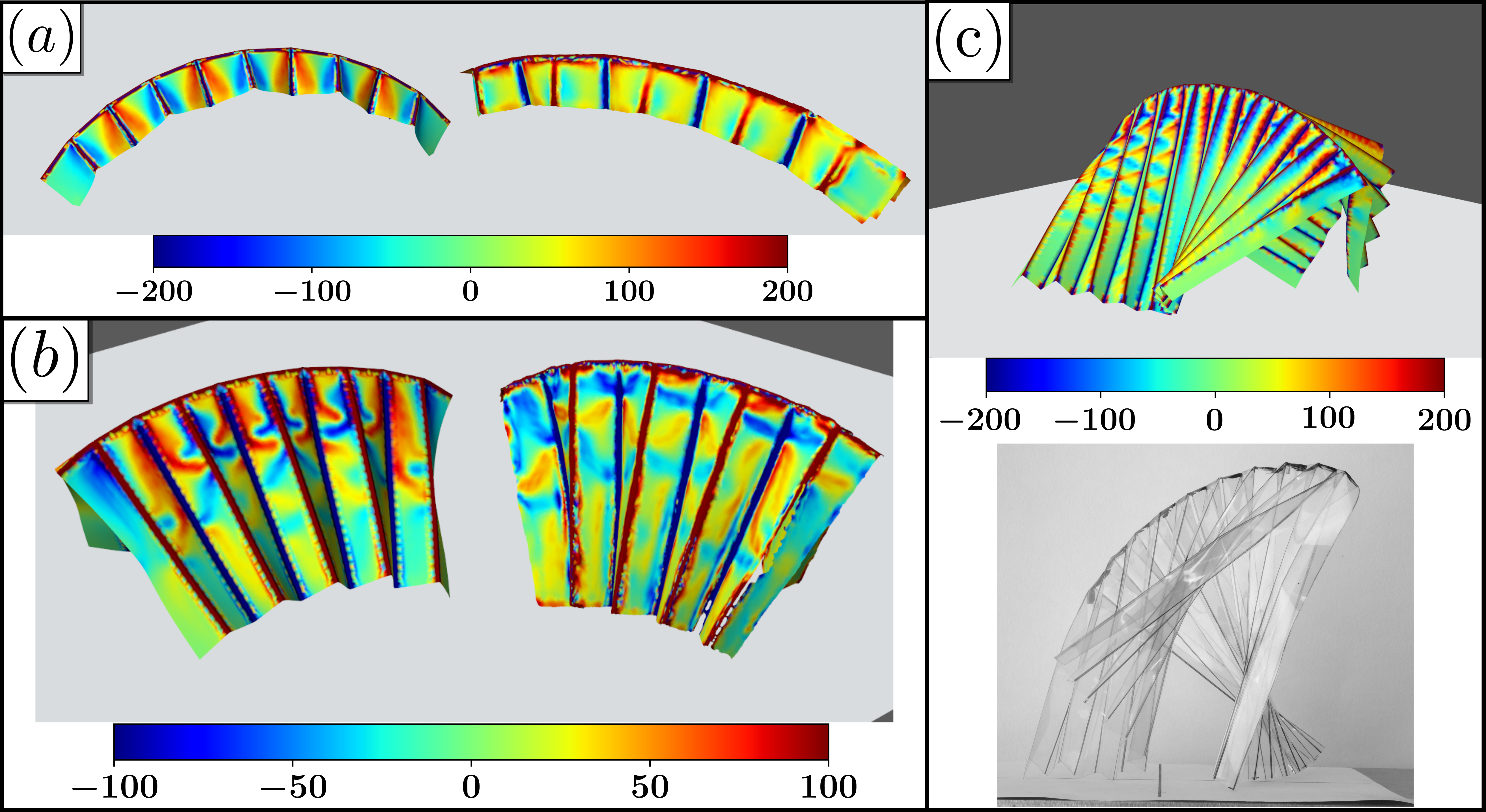}
\caption{Different elastic deformation regimes for the curved accordion origami. For all figures, the color represents the local mean curvature in m$^{-1}$. Typical deformations for the simulation (left/top) and the experiment (right/bottom) for (a) a single facet with $N=9$, $l = 20$~mm and $w=20$~mm, (b) faceting with $N=9$, $l = 100$~mm and $w=20$~mm, and (c) buckling with $N=19$, $l = 140$~mm and $w=10$~mm. The bottom image is a raw experimental image with $N=20$, $l=148.5$~mm and $w=10$~mm. The global shape is different from the simulation, however a buckling behavior is clearly observe with both methods. For buckled origami, 3D scanning is hard to perform due to the complex structure.}
\label{fig:DiffDef}
\end{center}
\end{figure}

Even with these discrepancies, both simulations and experiments display three typical deformation modes. A closer inspection of the numerical data in Fig.~(\ref{fig:RcEvol}) reveals additional structures underlying the evolution of the radius $R$ with respect to $l$ and $w$. These structures are confirmed when analyzing the dependency of $R$ on a single geometric parameter, for instance, the faces' length $l$. Fig.~\ref{fig:NonMonoRc} shows the existence of three regions $I$, $II$, and $III$, which correspond to very distinct regimes of deformation, as shown in Fig.~\ref{fig:DiffDef}. The region $I$ corresponds to a single facet deformation for which each face is separated into three areas. A central triangular flat facet is delimited by the central crease and two bent regions accommodating the stiffness of the side crease in a deformation very similar to the shape of elastic ridges~\cite{Witten2009}. Above a first critical length $l_1$, some faces begin to deform in two or more facets. This behavior points to the start of the region $II$, characterized by the tiling of the faces with triangular facets delimited by stretching ridges. The described deformation is similar to the one observed in twisted ribbons~\cite{Dinh2016}. A curving of the transverse creases themselves near the vertices accompany the faceting. Finally, above a second critical length $l_2$, the free edges initially parallel to the central crease buckle out of symmetry through Euler-like instability that defines the boundary of the region $III$.

\begin{figure}[htb]
\begin{center}
\includegraphics[width=\linewidth]{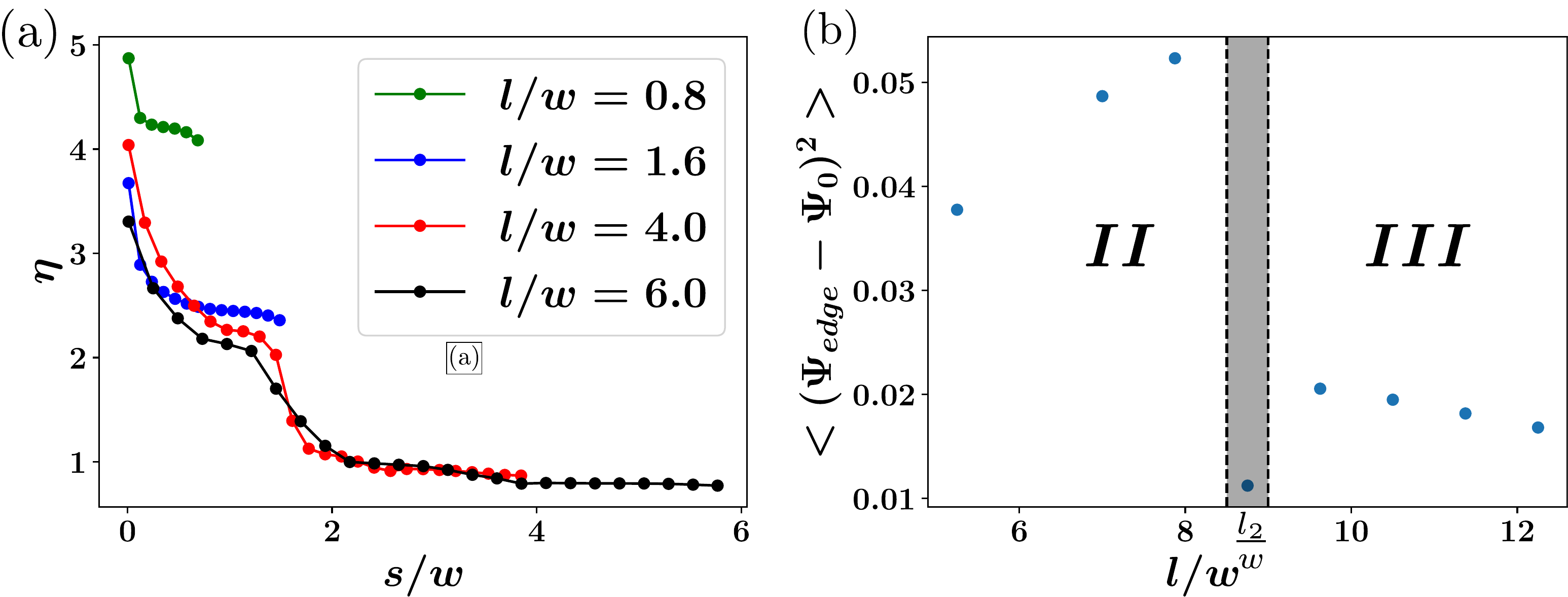}
\caption{(a) Local twist $\eta$ of a face next to the middle transverse crease, with $N=11$, $w=25$~mm, and different lengths $l$. $s=0$ at the central crease. (b) Squared angular deformation $(\Psi_{\text{edge}} - \Psi_0)^2$ averaged over all transverse creases for $N=13$ and $w=14.3$~mm. The vertical dotted line represent the approximate transition $l_2 \approx 9w$ between faceting and buckling. The point in the grey region display buckling only on one side of the central crease.}
\label{fig:TwistCompression}
\end{center}
\end{figure}

We drew ideas from mechanical systems with comparable geometry to characterize the transition between the deformation modes with the proper parameters. First, the faceting observed in region II is reminiscent of the shape displayed by twisted ribbons~\cite{Dinh2016, bohr2013ribbon, Chopin2019}, so we focused on quantifying the local twist in the faces. To that end, we extracted from the simulation the twist $\eta$ along a face next to the middle transverse crease $i=(N+1)/2$,
\begin{align}
\eta(s) \propto w\frac{\delta\phi}{\delta s}\;,
\label{eq:Twist}
\end{align}
where $s$ is the distance to the vertex along the transverse crease (see Fig.~\ref{fig:OrigamiShape}) and $\delta\phi$ is the angle after heating between two initially parallel lines at $s$ and $s+\delta s$. The twisting is constant along the whole face for single facet deformation. Strikingly, we notice from Fig.~\ref{fig:TwistCompression}(a) that $l_1$ not only indicates the transition to faceting but also defines the characteristic size of a boundary layer for high twist near the central crease. In this buffer region, the local deformation is different from the rest of the face. Beyond, the faces deform as twisted ribbons under slight tension~\cite{Chopin2016, Dinh2016} with a characteristic triangular faceting. The range over which the central crease affects the faces' deformations is of the order of $w$. The local influence of boundary conditions on the deformation of elastic origamis agrees with observations on their local actuation~\cite{Grey2019}.

Since the transverse creases remain mostly straight while faceting, the arc length at the free boundary is smaller than the arc of the central crease. As a result, the accordion is compressed near the free edge and store elastic crease energy, as shown in Fig.~\ref{fig:TwistCompression}~(b). Subsequently, for $l > l_2$, this energy becomes large, and the system prefers to adopt a new configuration that relaxes it by buckling similarly to compressed elastic rods. The buckling requires the faces and the transverse creases to bend and twist accordingly. Each side of the central crease has a different buckling threshold due to their different number of mountains and valleys folds. In our simulations, it results in a configuration for which only one side is buckling at the onset of the transition. This configuration displays a unique structure as highlighted in Fig.~\ref{fig:TwistCompression}~(b).

\begin{figure}[htb]
\begin{center}
\includegraphics[width=0.8\linewidth]{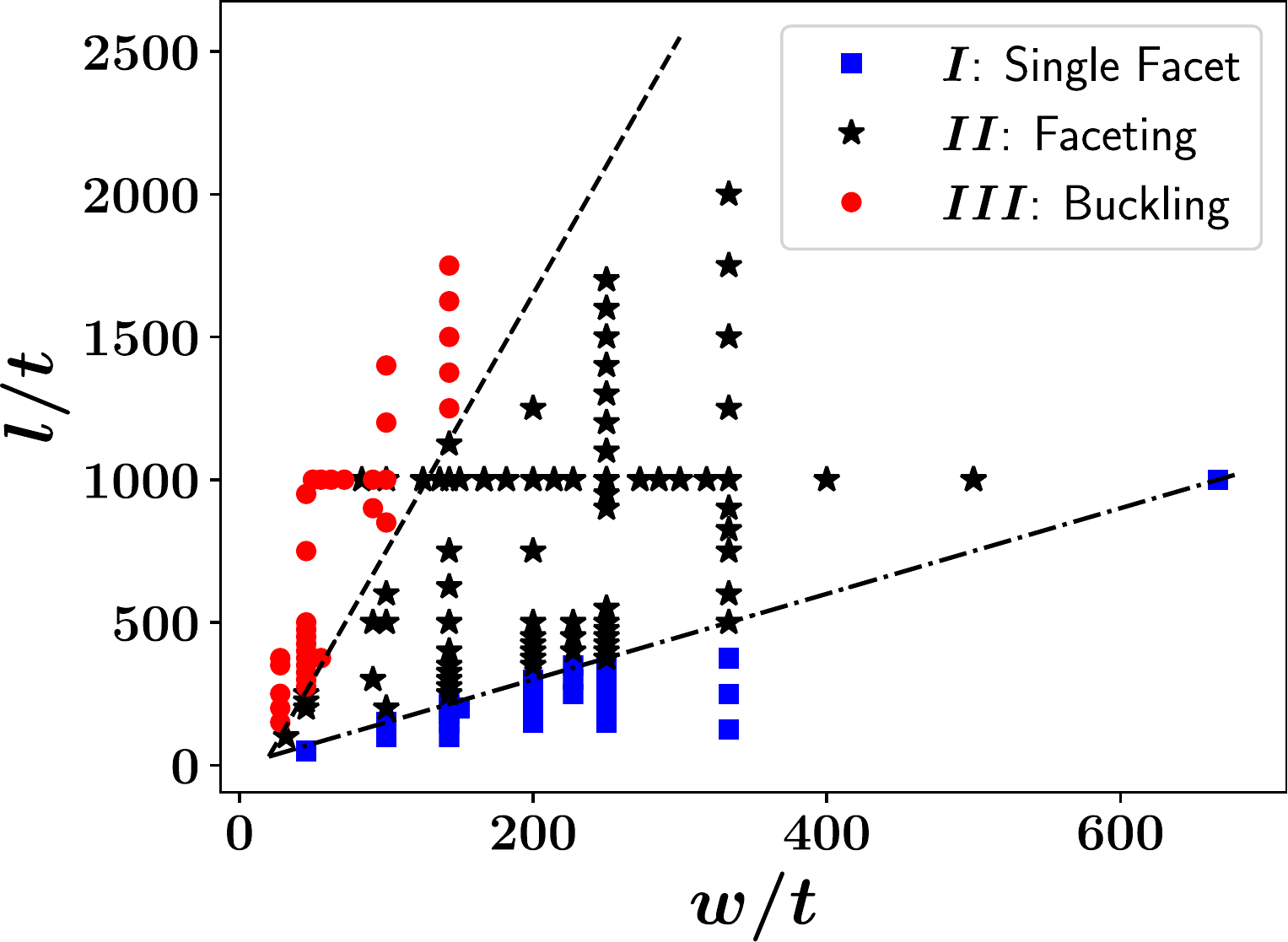}
\caption{Phase space of the regimes of deformations. The lines separating the different behaviors are given by $l_1 \approx 1.5w$ (dash-dotted) and $l_2 \approx 9w$ (dashed).}
\label{fig:PhaseSpace}
\end{center}
\end{figure}

With the three distinct modes of deformation we described, we draw in Fig.~\ref{fig:PhaseSpace} a phase diagram for the morphologies of the origami using the data from simulations for multiple patterns and dimensions. Just as we observed with the curvature, and even though the different elastic deformations rely on the origami's complex structures, the transitions between deformations only depend on the aspect ratio of the faces, with $l_1 \approx 1.5w$ and $l_2 \approx 9w$.

\section{Discussion}

Clearly, the geometry of the pattern governs the details of the shell-like shape of the accordion. This prominent effect comes from the stiffness of the creases. Since every accordion is stable, each shape corresponds to a minimal global elastic energy configuration that is the sum of contributions from the folds and the faces. Both depend on the elastic properties of the original sheet. However, in the limit of thin folds ($l,w\gg t$), the energy cost to open the crease ($E_{crease}\propto E l t^2$~\cite{Jules2019}) is high. Much more than the energy needed to deform the faces by bending ($E_{bending}\propto E t^3$~\cite{Audoly2010}) or, when the boundary conditions require non-isometric deformations, to produce stretching ridges ($E_{ridge}\propto E l^{1/3}t^{8/3}$~\cite{Witten2007}). Consequently, the faces' deformation is subordinate to the network of rigid creases, which acts as a backbone for the structure and imposes local boundary conditions. This characterization is evocative of elastic grid shells, where regular planar grids of thin elastic rods are actuated into three-dimensional shell-like structures by the loading of their extremities along a pre-determined path~\cite{Baek2019}. As a result, the accordion's effective curvature, characterized by $R$, is mainly governed by the pattern of folds and the value of the rest angles, while only marginally affected by the elasticity of the sheet. We verified this claim by varying the sheet's thickness by one order of magnitude (from \SIrange{20}{200}{\mu\meter}) in simulations and obtained very similar curvature  (See Supplemntary Materials).

\section{Conclusion}

Our study investigates a poorly developed branch of mechanical properties for origami, the frustrated shaping. Akin to shapes obtained with curved creases~\cite{Mouthuy2012, Dias2012a}, we manage to reach effective curvature for a deployed origami at equilibrium from a straightforward pattern of creases, a Miura folding with right angles. The interactions between stiff folds yield three global geometrical shapes for the deployed origami, resulting from the competition between various deformation modes of the faces such as bending, stretching, or twisting. 

Whereas we only considered the elasticity based on the material mechanical properties, the local nature of the competition shows a clear path to design by controlling the pattern of creases and the rest angle and stiffness of the folds. Contrary to rigid-faces origami design, the creases' network's choice does not limit the origami to a single configuration. The shape and the resulting properties are finely tunable after folding not only with local bistability~\cite{Silverberg2014}, but also through the precise changes of the rest angles~\cite{Grey2020}. However, quantitative applications will require a more in-depth elastic and geometric modeling of the complex competition between the creases and the faces' resulting deformations.

\section*{Acknowledgements}

T.-J. thanks Marcelo Dias and Dominic Vella for fruitful discussions and for providing access to COMSOL.

\begin{appendix}

\section{Supplementary Materials}

Supplementary Materials for further details on the 3D scanning method, the settings for the numerics and the effect of various parameters on the simulated curvature can be found in ...

\end{appendix}

\bibliographystyle{elsarticle-num.bst}
\bibliography{BiblioAcc}

\begin{thebibliography}{10}
\expandafter\ifx\csname url\endcsname\relax
  \def\url#1{\texttt{#1}}\fi
\expandafter\ifx\csname urlprefix\endcsname\relax\def\urlprefix{URL }\fi
\expandafter\ifx\csname href\endcsname\relax
  \def\href#1#2{#2} \def\path#1{#1}\fi

\bibitem{Lv2014}
C.~Lv, D.~Krishnaraju, G.~Konjevod, H.~Yu, H.~Jiang, Origami based mechanical
  metamaterials, Scientific Reports 4~(1) (aug 2014).
\newblock \href {https://doi.org/10.1038/srep05979}
  {\path{doi:10.1038/srep05979}}.

\bibitem{Silverberg2014}
J.~L. Silverberg, A.~A. Evans, L.~McLeod, R.~C. Hayward, T.~Hull, C.~D.
  Santangelo, I.~Cohen, Using origami design principles to fold reprogrammable
  mechanical metamaterials, Science 345~(6197) (2014) 647--650.
\newblock \href {https://doi.org/10.1126/science.1252876}
  {\path{doi:10.1126/science.1252876}}.

\bibitem{Overvelde2016}
J.~T. Overvelde, T.~A. de~Jong, Y.~Shevchenko, S.~A. Becerra, G.~M. Whitesides,
  J.~C. Weaver, C.~Hoberman, K.~Bertoldi, A three-dimensional actuated
  origami-inspired transformable metamaterial with multiple degrees of freedom,
  Nature Communications 7~(1) (mar 2016).
\newblock \href {https://doi.org/10.1038/ncomms10929}
  {\path{doi:10.1038/ncomms10929}}.

\bibitem{Tachi2010}
T.~Tachi, Geometric considerations for the design of rigid origami structures,
  in: Proceedings of the International Association for Shell and Spatial
  Structures (IASS) Symposium, Elsevier Ltd, 2010, pp. 458--460.

\bibitem{Schenk2013}
M.~Schenk, S.~D. Guest, Geometry of miura-folded metamaterials, Proceedings of
  the National Academy of Sciences 110~(9) (2013) 3276--3281.
\newblock \href {https://doi.org/10.1073/pnas.1217998110}
  {\path{doi:10.1073/pnas.1217998110}}.

\bibitem{Wei2013}
Z.~Y. Wei, Z.~V. Guo, L.~Dudte, H.~Y. Liang, L.~Mahadevan, Geometric mechanics
  of periodic pleated origami, Physical Review Letters 110~(21) (may 2013).
\newblock \href {https://doi.org/10.1103/physrevlett.110.215501}
  {\path{doi:10.1103/physrevlett.110.215501}}.

\bibitem{Pratapa2019}
P.~P. Pratapa, K.~Liu, G.~H. Paulino, Geometric mechanics of origami patterns
  exhibiting poisson's ratio switch by breaking mountain and valley assignment,
  Physical Review Letters 122~(15) (apr 2019).
\newblock \href {https://doi.org/10.1103/physrevlett.122.155501}
  {\path{doi:10.1103/physrevlett.122.155501}}.

\bibitem{Dudte2016}
L.~H. Dudte, E.~Vouga, T.~Tachi, L.~Mahadevan, Programming curvature using
  origami~tessellations, Nature Materials 15~(5) (2016) 583--588.
\newblock \href {https://doi.org/10.1038/nmat4540}
  {\path{doi:10.1038/nmat4540}}.

\bibitem{Nassar2017}
H.~Nassar, A.~Leb{\'{e}}e, L.~Monasse, Curvature, metric and parametrization of
  origami tessellations: theory and application to the eggbox pattern,
  Proceedings of the Royal Society A: Mathematical, Physical and Engineering
  Sciences 473~(2197) (2017) 20160705.
\newblock \href {https://doi.org/10.1098/rspa.2016.0705}
  {\path{doi:10.1098/rspa.2016.0705}}.

\bibitem{Hayakawa2020}
K.~Hayakawa, M.~Ohsaki, Form generation of rigid origami for approximation of a
  curved surface based on mechanical property of partially rigid frames,
  International Journal of Solids and Structures (dec 2020).
\newblock \href {https://doi.org/10.1016/j.ijsolstr.2020.12.007}
  {\path{doi:10.1016/j.ijsolstr.2020.12.007}}.

\bibitem{Reid2017}
A.~Reid, F.~Lechenault, S.~Rica, M.~Adda-Bedia, Geometry and design of origami
  bellows with tunable response, Physical Review E 95~(1) (jan 2017).
\newblock \href {https://doi.org/10.1103/physreve.95.013002}
  {\path{doi:10.1103/physreve.95.013002}}.

\bibitem{Silverberg2015}
J.~L. Silverberg, J.-H. Na, A.~A. Evans, B.~Liu, T.~C. Hull, C.~D. Santangelo,
  R.~J. Lang, R.~C. Hayward, I.~Cohen, Origami structures with a critical
  transition to bistability arising from hidden degrees of freedom, Nature
  Materials 14~(4) (2015) 389--393.
\newblock \href {https://doi.org/10.1038/nmat4232}
  {\path{doi:10.1038/nmat4232}}.

\bibitem{Sultan2006}
E.~Sultan, A.~Boudaoud, Statistics of crumpled paper, Physical Review Letters
  96~(13) (apr 2006).
\newblock \href {https://doi.org/10.1103/physrevlett.96.136103}
  {\path{doi:10.1103/physrevlett.96.136103}}.

\bibitem{Gottesman2018}
O.~Gottesman, J.~Andrejevic, C.~H. Rycroft, S.~M. Rubinstein, A state variable
  for crumpled thin sheets, Communications Physics 1~(1) (nov 2018).
\newblock \href {https://doi.org/10.1038/s42005-018-0072-x}
  {\path{doi:10.1038/s42005-018-0072-x}}.

\bibitem{Amar1997}
M.~B. Amar, Y.~Pomeau, Crumpled paper, Proceedings of the Royal Society A:
  Mathematical, Physical and Engineering Sciences 453~(1959) (1997) 729--755.
\newblock \href {https://doi.org/10.1098/rspa.1997.0041}
  {\path{doi:10.1098/rspa.1997.0041}}.

\bibitem{Walker2018}
M.~G. Walker, K.~A. Seffen, On the shape of bistable creased strips,
  Thin-Walled Structures 124 (2018) 538--545.
\newblock \href {https://doi.org/10.1016/j.tws.2017.12.033}
  {\path{doi:10.1016/j.tws.2017.12.033}}.

\bibitem{Andrade-Silva2019}
I.~Andrade-Silva, M.~Adda-Bedia, M.~A. Dias, Foldable cones as a framework for
  nonrigid origami, Physical Review E 100~(3) (sep 2019).
\newblock \href {https://doi.org/10.1103/physreve.100.033003}
  {\path{doi:10.1103/physreve.100.033003}}.

\bibitem{Demaine2011}
E.~D. Demaine, M.~L. Demaine, D.~Koschitz, T.~Tachi, Curved crease folding: a
  review on art, design and mathematics, in: Proceedings of the IABSE-IASS
  Symposium: Taller, Longer, Lighter, IABSE-IASS, 2011, pp. 20--23.

\bibitem{Dias2012}
M.~A. Dias, L.~H. Dudte, L.~Mahadevan, C.~D. Santangelo, Geometric mechanics of
  curved crease origami, Physical Review Letters 109~(11) (sep 2012).
\newblock \href {https://doi.org/10.1103/physrevlett.109.114301}
  {\path{doi:10.1103/physrevlett.109.114301}}.

\bibitem{Mouthuy2012}
P.-O. Mouthuy, M.~Coulombier, T.~Pardoen, J.-P. Raskin, A.~M. Jonas,
  Overcurvature describes the buckling and folding of rings from curved origami
  to foldable tents, Nature Communications 3~(1) (jan 2012).
\newblock \href {https://doi.org/10.1038/ncomms2311}
  {\path{doi:10.1038/ncomms2311}}.

\bibitem{Tachi2013a}
T.~Tachi, Designing freeform origami tessellations by generalizing
  resch{\textquotesingle}s patterns, Journal of Mechanical Design 135~(11) (oct
  2013).
\newblock \href {https://doi.org/10.1115/1.4025389}
  {\path{doi:10.1115/1.4025389}}.

\bibitem{Chen2016}
Y.~Chen, H.~Feng, J.~Ma, R.~Peng, Z.~You, Symmetric waterbomb origami,
  Proceedings of the Royal Society A: Mathematical, Physical and Engineering
  Sciences 472~(2190) (2016) 20150846.
\newblock \href {https://doi.org/10.1098/rspa.2015.0846}
  {\path{doi:10.1098/rspa.2015.0846}}.

\bibitem{Baek2019}
C.~Baek, P.~M. Reis, Rigidity of hemispherical elastic gridshells under point
  load indentation, Journal of the Mechanics and Physics of Solids 124 (2019)
  411--426.
\newblock \href {https://doi.org/10.1016/j.jmps.2018.11.002}
  {\path{doi:10.1016/j.jmps.2018.11.002}}.

\bibitem{Grey2020}
S.~W. Grey, F.~Scarpa, M.~Schenk, Mechanics of paper-folded origami: A
  cautionary tale, Mechanics Research Communications 107 (2020) 103540.
\newblock \href {https://doi.org/10.1016/j.mechrescom.2020.103540}
  {\path{doi:10.1016/j.mechrescom.2020.103540}}.

\bibitem{Jules2019}
T.~Jules, F.~Lechenault, M.~Adda-Bedia, Local mechanical description of an
  elastic fold, Soft Matter 15~(7) (2019) 1619--1626.
\newblock \href {https://doi.org/10.1039/c8sm01791c}
  {\path{doi:10.1039/c8sm01791c}}.

\bibitem{Lazarus2012}
A.~Lazarus, H.~C.~B. Florijn, P.~M. Reis, Geometry-induced rigidity in
  nonspherical pressurized elastic shells, Physical Review Letters 109~(14)
  (oct 2012).
\newblock \href {https://doi.org/10.1103/physrevlett.109.144301}
  {\path{doi:10.1103/physrevlett.109.144301}}.

\bibitem{Na2014}
J.-H. Na, A.~A. Evans, J.~Bae, M.~C. Chiappelli, C.~D. Santangelo, R.~J. Lang,
  T.~C. Hull, R.~C. Hayward, Programming reversibly self-folding origami with
  micropatterned photo-crosslinkable polymer trilayers, Advanced Materials
  27~(1) (2014) 79--85.
\newblock \href {https://doi.org/10.1002/adma.201403510}
  {\path{doi:10.1002/adma.201403510}}.

\bibitem{Benusiglio2012}
A.~Benusiglio, V.~Mansard, A.-L. Biance, L.~Bocquet, The anatomy of a crease,
  from folding to ironing, Soft Matter 8~(12) (2012) 3342.
\newblock \href {https://doi.org/10.1039/c2sm07151g}
  {\path{doi:10.1039/c2sm07151g}}.

\bibitem{Lechenault2015}
F.~Lechenault, M.~Adda-Bedia, Generic bistability in creased conical surfaces,
  Physical Review Letters 115~(23) (dec 2015).
\newblock \href {https://doi.org/10.1103/physrevlett.115.235501}
  {\path{doi:10.1103/physrevlett.115.235501}}.

\bibitem{Witten2009}
T.~A. Witten, Spontaneous free-boundary structure in crumpled
  membranes{\textdagger}, The Journal of Physical Chemistry B 113~(12) (2009)
  3738--3742.
\newblock \href {https://doi.org/10.1021/jp807548s}
  {\path{doi:10.1021/jp807548s}}.

\bibitem{Dinh2016}
H.~P. Dinh, V.~D{\'{e}}mery, B.~Davidovitch, F.~Brau, P.~Damman, From
  cylindrical to stretching ridges and wrinkles in twisted ribbons, Physical
  Review Letters 117~(10) (sep 2016).
\newblock \href {https://doi.org/10.1103/physrevlett.117.104301}
  {\path{doi:10.1103/physrevlett.117.104301}}.

\bibitem{bohr2013ribbon}
J.~Bohr, S.~Markvorsen, Ribbon crystals, Plos one 8~(10) (2013) e74932.

\bibitem{Chopin2019}
J.~Chopin, R.~T.~D. Filho, Extreme contractility and torsional compliance of
  soft ribbons under high twist, Physical Review E 99~(4) (2019) 043002.
\newblock \href {https://doi.org/10.1103/physreve.99.043002}
  {\path{doi:10.1103/physreve.99.043002}}.

\bibitem{Chopin2016}
J.~Chopin, V.~D{\'{e}}mery, B.~Davidovitch, Roadmap to the morphological
  instabilities of a stretched twisted ribbon, in: The Mechanics of Ribbons and
  Möbius Bands, Springer Netherlands, 2016, pp. 137--189.
\newblock \href {https://doi.org/10.1007/978-94-017-7300-3_10}
  {\path{doi:10.1007/978-94-017-7300-3_10}}.

\bibitem{Grey2019}
S.~W. Grey, F.~Scarpa, M.~Schenk, Strain reversal in actuated origami
  structures, Physical Review Letters 123~(2) (jul 2019).
\newblock \href {https://doi.org/10.1103/physrevlett.123.025501}
  {\path{doi:10.1103/physrevlett.123.025501}}.

\bibitem{Audoly2010}
B.~Audoly, Y.~Pomeau, Elasticity and geometry: from hair curls to the
  non-linear response of shells, Oxford University Press, 2010.

\bibitem{Witten2007}
T.~A. Witten, Stress focusing in elastic sheets, Reviews of Modern Physics
  79~(2) (2007) 643--675.
\newblock \href {https://doi.org/10.1103/revmodphys.79.643}
  {\path{doi:10.1103/revmodphys.79.643}}.

\bibitem{Dias2012a}
M.~A. Dias, C.~D. Santangelo, The shape and mechanics of curved-fold origami
  structures, {EPL} (Europhysics Letters) 100~(5) (2012) 54005.
\newblock \href {https://doi.org/10.1209/0295-5075/100/54005}
  {\path{doi:10.1209/0295-5075/100/54005}}.

\end{thebibliography}


\begin{thebibliography}{3}%
\makeatletter
\providecommand \@ifxundefined [1]{%
 \@ifx{#1\undefined}
}%
\providecommand \@ifnum [1]{%
 \ifnum #1\expandafter \@firstoftwo
 \else \expandafter \@secondoftwo
 \fi
}%
\providecommand \@ifx [1]{%
 \ifx #1\expandafter \@firstoftwo
 \else \expandafter \@secondoftwo
 \fi
}%
\providecommand \natexlab [1]{#1}%
\providecommand \enquote  [1]{``#1''}%
\providecommand \bibnamefont  [1]{#1}%
\providecommand \bibfnamefont [1]{#1}%
\providecommand \citenamefont [1]{#1}%
\providecommand \href@noop [0]{\@secondoftwo}%
\providecommand \href [0]{\begingroup \@sanitize@url \@href}%
\providecommand \@href[1]{\@@startlink{#1}\@@href}%
\providecommand \@@href[1]{\endgroup#1\@@endlink}%
\providecommand \@sanitize@url [0]{\catcode `\\12\catcode `\$12\catcode
  `\&12\catcode `\#12\catcode `\^12\catcode `\_12\catcode `\%12\relax}%
\providecommand \@@startlink[1]{}%
\providecommand \@@endlink[0]{}%
\providecommand \url  [0]{\begingroup\@sanitize@url \@url }%
\providecommand \@url [1]{\endgroup\@href {#1}{\urlprefix }}%
\providecommand \urlprefix  [0]{URL }%
\providecommand \Eprint [0]{\href }%
\providecommand \doibase [0]{https://doi.org/}%
\providecommand \selectlanguage [0]{\@gobble}%
\providecommand \bibinfo  [0]{\@secondoftwo}%
\providecommand \bibfield  [0]{\@secondoftwo}%
\providecommand \translation [1]{[#1]}%
\providecommand \BibitemOpen [0]{}%
\providecommand \bibitemStop [0]{}%
\providecommand \bibitemNoStop [0]{.\EOS\space}%
\providecommand \EOS [0]{\spacefactor3000\relax}%
\providecommand \BibitemShut  [1]{\csname bibitem#1\endcsname}%
\let\auto@bib@innerbib\@empty
\bibitem [{\citenamefont {Cignoni}\ \emph {et~al.}(2008)\citenamefont
  {Cignoni}, \citenamefont {Callieri}, \citenamefont {Corsini}, \citenamefont
  {Dellepiane}, \citenamefont {Ganovelli},\ and\ \citenamefont
  {Ranzuglia}}]{MeshLab}%
  \BibitemOpen
  \bibfield  {author} {\bibinfo {author} {\bibfnamefont {P.}~\bibnamefont
  {Cignoni}}, \bibinfo {author} {\bibfnamefont {M.}~\bibnamefont {Callieri}},
  \bibinfo {author} {\bibfnamefont {M.}~\bibnamefont {Corsini}}, \bibinfo
  {author} {\bibfnamefont {M.}~\bibnamefont {Dellepiane}}, \bibinfo {author}
  {\bibfnamefont {F.}~\bibnamefont {Ganovelli}},\ and\ \bibinfo {author}
  {\bibfnamefont {G.}~\bibnamefont {Ranzuglia}},\ }\bibfield  {title} {\bibinfo
  {title} {Meshlab: an open-source mesh processing tool},\ }in\ \href
  {https://doi.org/10.2312/LocalChapterEvents/ItalChap/ItalianChapConf2008/129-136}
  {\emph {\bibinfo {booktitle} {Eurographics Italian Chapter Conference}}},\
  \bibinfo {editor} {edited by\ \bibinfo {editor} {\bibfnamefont
  {V.}~\bibnamefont {Scarano}}, \bibinfo {editor} {\bibfnamefont {R.~D.}\
  \bibnamefont {Chiara}},\ and\ \bibinfo {editor} {\bibfnamefont
  {U.}~\bibnamefont {Erra}}}\ (\bibinfo  {publisher} {The Eurographics
  Association},\ \bibinfo {year} {2008})\BibitemShut {NoStop}%
\bibitem [{\citenamefont {Guennebaud}\ and\ \citenamefont
  {Gross}(2007)}]{Guennebaud2007}%
  \BibitemOpen
  \bibfield  {author} {\bibinfo {author} {\bibfnamefont {G.}~\bibnamefont
  {Guennebaud}}\ and\ \bibinfo {author} {\bibfnamefont {M.}~\bibnamefont
  {Gross}},\ }\bibfield  {title} {\bibinfo {title} {Algebraic point set
  surfaces},\ }in\ \href {https://doi.org/10.1145/1275808.1276406} {\emph
  {\bibinfo {booktitle} {{ACM} {SIGGRAPH} 2007 papers on - {SIGGRAPH}
  {\textquotesingle}07}}}\ (\bibinfo  {publisher} {{ACM} Press},\ \bibinfo
  {year} {2007})\BibitemShut {NoStop}%
\bibitem [{\citenamefont {Guennebaud}\ \emph {et~al.}(2008)\citenamefont
  {Guennebaud}, \citenamefont {Germann},\ and\ \citenamefont
  {Gross}}]{Guennebaud2008}%
  \BibitemOpen
  \bibfield  {author} {\bibinfo {author} {\bibfnamefont {G.}~\bibnamefont
  {Guennebaud}}, \bibinfo {author} {\bibfnamefont {M.}~\bibnamefont
  {Germann}},\ and\ \bibinfo {author} {\bibfnamefont {M.}~\bibnamefont
  {Gross}},\ }\bibfield  {title} {\bibinfo {title} {Dynamic sampling and
  rendering of algebraic point set surfaces},\ }\href
  {https://doi.org/10.1111/j.1467-8659.2008.01163.x} {\bibfield  {journal}
  {\bibinfo  {journal} {Computer Graphics Forum}\ }\textbf {\bibinfo {volume}
  {27}},\ \bibinfo {pages} {653} (\bibinfo {year} {2008})}\BibitemShut
  {NoStop}%
\end{thebibliography}%

\end{document}


\title{Curving Origami with Mechanical Frustration \\ Supplementary Materials}

\begin{abstract}
\end{abstract}

\maketitle

\section*{3D Scanning and analysis.}

Structured light scanning is a technique to measure an object's three-dimensional shape using a projected light pattern and a camera. For our experiment, we use the DAVID 3D software coupled with a webcam Logitech C920 Pro HD and a projector ACER H6517ST. After a calibration step achieved by the software on a specifically patterned right angle, we analyze the shape of the origamis under multiple orientations thanks to a rotating platform. To ensure the origami is visible, we fold the transparent elastic sheet and then spray opaque white paint on top of the external surface. Then, we fuse the multiple meshes we obtain into a single one and analyze it.

To get the local curvature from the mesh, we use the software MeshLab~\cite{MeshLab}, and more precisely, the filter Colorize curvature (APSS)~\cite{Guennebaud2007, Guennebaud2008} and changing the MLS - Filter scale to 10 in order to reduce the noise of the scan. The other parameters are kept to their default values. 

\section*{Folding simulation}

\subsection*{Folding simulation}

To simulate the folding of the curved accordion, we begin by setting the geometry of the pattern. To do that, we start by modelling a rectangular surface of dimensions $2l \times (N+1)w$. Then, we delimit the pattern of creases on the surface. Every crease is centered on the pattern and is $2S_0$ wide, with $S_0 = 3t$, where $t$ is the thickness of the sheet. We puncture a hole of radius $\sqrt{2}S_0$ at every vertex to avoid a too large stress concentration. Finally, we only choose an odd number of creases to have a symmetric crease pattern and thus to simulate only half of the system. Thanks to each side of this crease's symmetry, we increase the computing speed.

\begin{figure}[h]
\begin{center}
\includegraphics[width=0.8\linewidth]{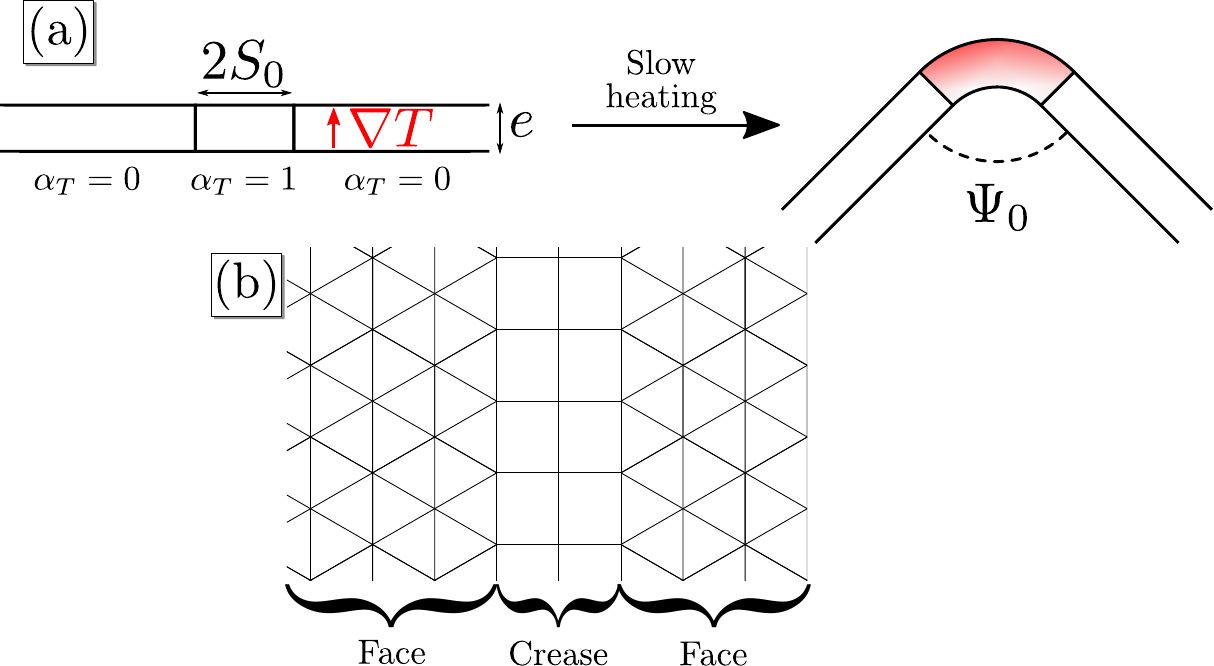}\\
\caption{(a) Schematics of the folding through thermal dilatation of a crease of size $2S_0$. (b) Typical mesh around the crease.}
\label{fig:CreationCrease}
\end{center}
\end{figure}

For the meshing, we differentiate between the faces and the creases. For the former, we choose free triangular meshes with a size \emph{fine} to \emph{extremely fine}. For the latter, we choose rectangular meshes that ensure a line of nodes along the center line of the creases while connecting the vertex of triangles from the two connected faces with a straight line. A schematics of it is shown in Fig.~\ref{fig:CreationCrease} (b).

To incorporate the physics, we choose to model our system with the \emph{shell} model from COMSOL, suited for analyzing 3D structures created from thin sheets. With this model, we only need to produce a 2D surface, the thickness being added by the equations of the model afterward. Then, we impose a thermal dilatation coefficient $\alpha_T$ to each crease depending on whether it is a mountain ($\alpha_T = +1$ K$^{-1}$) or a valley ($\alpha_T = -1$ K$^{-1}$). At this step, we also attribute to the sheet its mechanical properties, its Young modulus $E=4$ GPa, and its Poisson ratio $\nu = 0.4$.

To fold the crease, we impose a gradient of temperature along with the thickness of the creases. While the temperature at the bottom does not change, the one at the top increases. The resulting dilatation of the crease folds it, as schematized in Fig.~\ref{fig:CreationCrease} (a).

The folding is done step by step, slowly increasing the gradient until it reaches a temperature $\Delta T$. We also choose to separate the heating of the central crease and the transverse creases. The heating protocol is as follow:
We start by heating the central crease until the gradient reaches $0.2 \Delta T$. Then, we heat the transverse creases until the gradient also reaches $0.2 \Delta T$. Finally, we repeat the operation by $0.2 \Delta T$ increment until all the creases reach a gradient of $\Delta T$.

We set three kinematic constraints for the accordion, all at the $(N+1)/2$ transverse crease. First, the center line of the crease is set as a symmetry for the whole system. On this line, we add constraints to the point at each extremity of this line, at the free boundaries. We prescribe a null displacement to one, while the other is only allowed to move on the line defined by these two points.

We choose the temperature $\Delta T$ such that a target rest angle $\Psi_0$ is reached. The dilation of the top part of the crease increases its spatial width up to
\begin{align}
L_t = (1+\alpha_T \Delta T)2S_0.
\label{eq:DilationCrease}
\end{align}
Through simple geometrical relations, $L_t$ is linked to the angular opening of the crease at rest by
\begin{align}
\pi - \Psi_0 &= \frac{L_h-2S_o}{t}\\
&= 2\alpha_T \Delta T \frac{S_o}{t}.
\label{eq:AngleTempTh}
\end{align}
We verified Eq.~(\ref{eq:AngleTempTh}) by simulating a single fold for multiple thickness $t$ and size of crease $S_0$. We found a similar relation
\begin{align}
\pi - \Psi_0 = A\alpha_T \Delta T \frac{S_o}{t},
\label{eq:AngleTempSimu}
\end{align}
with $A=2.8$. To reach a rest angle of 90 deg, we set the temperature to $\Delta T = 0.187$ K.

A large part of the resolution method we used is set by default by COMSOL. We use the MUMPS algorithm with a pivot threshold of 0.1 and a relative tolerance of 0.001. Two parameters we change are the maximum number of iterations that we set to 250 to let the algorithm more time to reach the solution, and we take into account the geometrical nonlinearities.

\subsection*{Effect of the rest angle}

\begin{figure}[h]
\begin{center}
\includegraphics[width=0.4\linewidth]{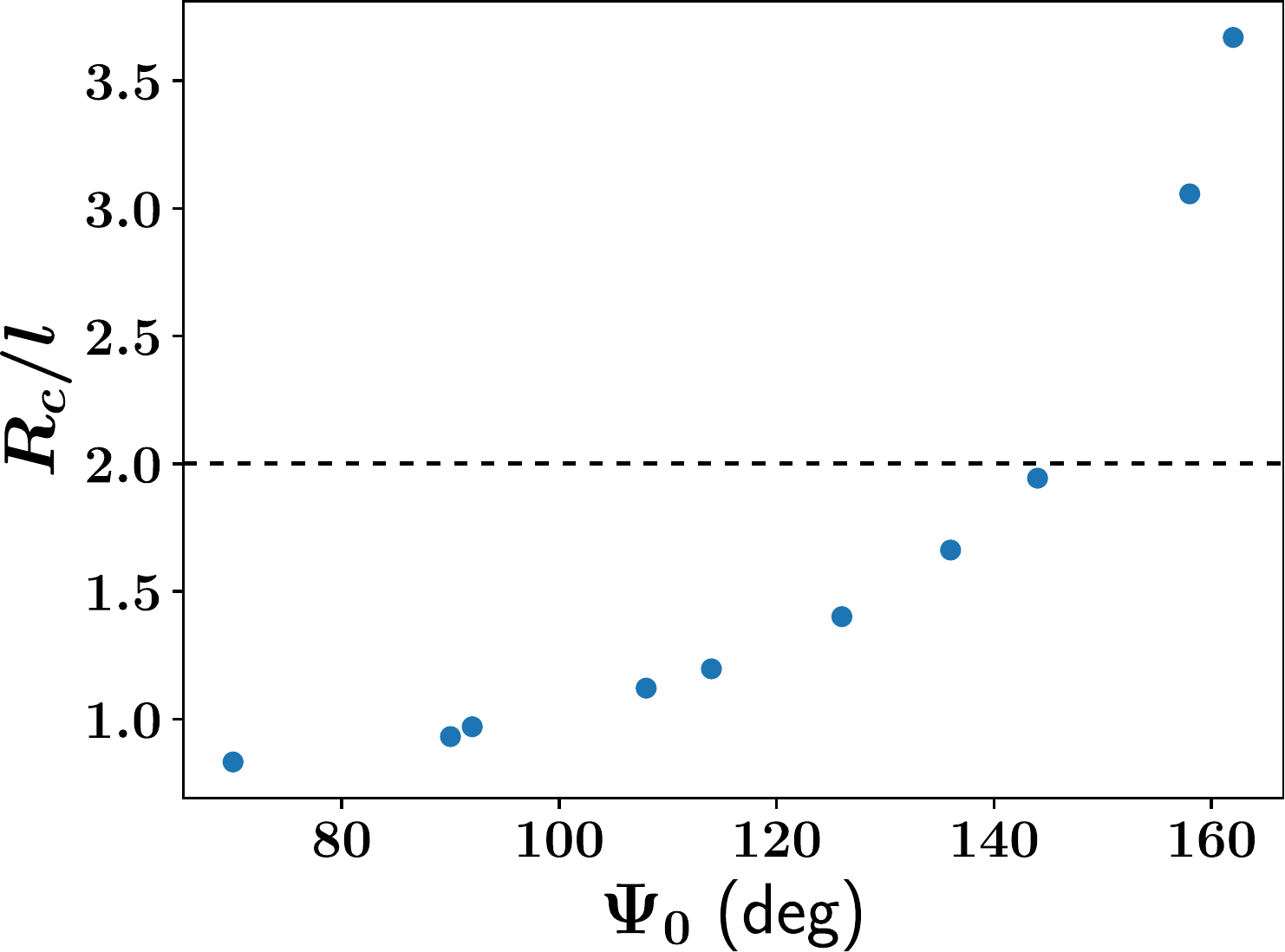}\\
\caption{Evolution of normalized radius of curvature $R_c/l$ with respect to
the rest angle of every crease $\Psi_0$, for an accordion of parameters $N=13$, $l=150$ mm and $w=14.3$ mm. The dotted line represent the curvature found in the experiment around this aspect ratio.}
\label{fig:RestAngle}
\end{center}
\end{figure}

As we detail in the previous section, we increase the heating of the central crease and of all the transverse creases alternatively during the folding process. As a result, there are multiple solutions for which every crease's rest angle is the same. In Fig.~\ref{fig:RestAngle}, we plot the evolution of the corresponding curvature of the accordion with respect to the rest angle.

We observe a clear nonlinear increase of $R_c$ when we open the rest angle. As expected, $R_c \rightarrow \infty$ when $\Psi_0 \rightarrow 180^\degree$ since it get closer and closer to the flat sheet. We also notice that the experimental data hint at an effective rest angle much larger than 90$^\degree$. A part of it may come from the experimental protocol since we do not control precisely the rest angles, which could be higher than 90$^\degree$. However, the angle 145$^\degree$ expected from the simulation seems too large for the shape we observe experimentally. So the difference of curvature between experiments and simulations has also another origin.

\subsection*{Effect of the sheet thickness and the holes at the vertices}

\begin{figure}[h]
\begin{center}
\includegraphics[width=\linewidth]{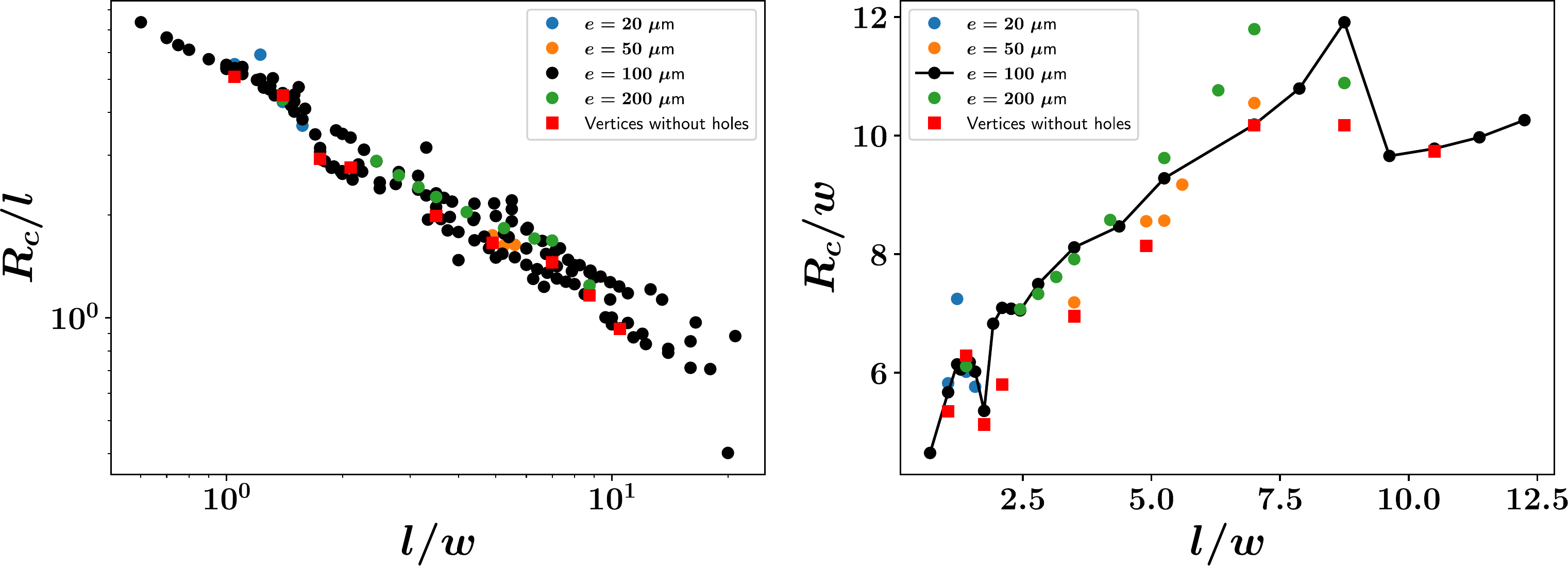}\\
\caption{(a) Evolution of normalized radius $R_c/l$ with respect to
the aspect ratio $l/w$. (b) Evolution of normalized radius $R_c/l$ with respect to the aspect ratio $l/w$ with $N=13$ and $w=14.3$ mm. For both figures the red square correspond to 100 $\mu$m thick sheet without hole at the vertices and the color of the circle correspond to different thickness from 20 to 200 $\mu$m.  }
\label{fig:ThicknessHoles}
\end{center}
\end{figure}

An essential assertion in this work is that the overall shape mainly depends on the pattern of creases and not the sheet's elasticity, as long as the thickness is negligible compared to the origami's characteristic lengths. To confirm our claim, we simulate folding the curved accordion for identical patterns but multiple thicknesses of sheets, from 20 to 200 $\mu$m, one order of magnitude, for $N=13$ and $w=14.3$ mm. When compared with previous data for a thickness of 100 $\mu$m in Fig.~\ref{fig:ThicknessHoles}, we notice that the thickness only has a marginal impact on the global curvature of the accordion.

A final element to analyze is the impact of the holes at the vertices used in our simulations to minimize the solving issues caused by the stress concentration at the vertices. Since they represent a sizeable structural difference between experiments and simulations, they might be responsible for the inadequacy between the curvature of both types of data. However, we see in Fig.~\ref{fig:ThicknessHoles} that the holes only have a minimal impact on the resulting curvature.

\bibliography{BiblioSM}